\documentclass[sigconf, authorversion]{acmart}

\usepackage{lipsum}
\usepackage{subcaption}
\usepackage{macros}
\usepackage{amsmath}
\usepackage{multirow}
\usepackage{subcaption}
\usepackage{wrapfig}
\usepackage{epigraph}
\usepackage{graphics}

\AtBeginDocument{%
  \providecommand\BibTeX{{%
    \normalfont B\kern-0.5em{\scshape i\kern-0.25em b}\kern-0.8em\TeX}}}

\setcopyright{rightsretained}

\begin{document}

\title[Cheat Sheet for Teaching Programming with Comics]{Cheat Sheet for Teaching Programming with Comics: \\Through the Lens of Concept-Language-Procedure Framework}

\author{Sangho Suh}
\orcid{0000-0003-4617-5116}
\affiliation{%
  \institution{University of Waterloo}
  \city{Waterloo}
  \country{Canada}
}
\email{sangho.suh@uwaterloo.ca}


\renewcommand{\shortauthors}{Sangho Suh}

\begin{abstract}
Comics is emerging as a popular medium for providing visual explanations of programming concepts and procedures. Recent research into this medium opened the door to new opportunities and tools to advance teaching and learning in computing. For instance, recent research on coding strip, a form of comic strip with its corresponding code, led to a new visual programming environment that generates comics from code and experience report detailing various ways coding strips can be used to benefit students' learning. However, how comics can be designed and used to teach programming has not yet been documented in a concise, accessible format to ease their adoption. To fill this gap, we developed a cheat sheet that summarizes the pedagogical techniques and designs teachers can use in their teaching. To develop this cheat sheet, we analyzed prior work on coding strip, including 26 coding strips and 30 coding strip design patterns. We also formulated a concept-language-procedure framework to delineate how comics can facilitate teaching in programming. To evaluate our cheat sheet, we presented it to 11 high school CS teachers at an annual conference for computer studies educators and asked them to rate its readability, usefulness, organization, and their interest in using it for their teaching. Our analysis suggests that this cheat sheet is easy to read/understand, useful, well-structured, and interests teachers to further explore how they can incorporate comics into their teaching.
\end{abstract}



\begin{teaserfigure}
  \includegraphics[trim=0cm 0cm 0cm 0cm, clip=true, width=\textwidth]{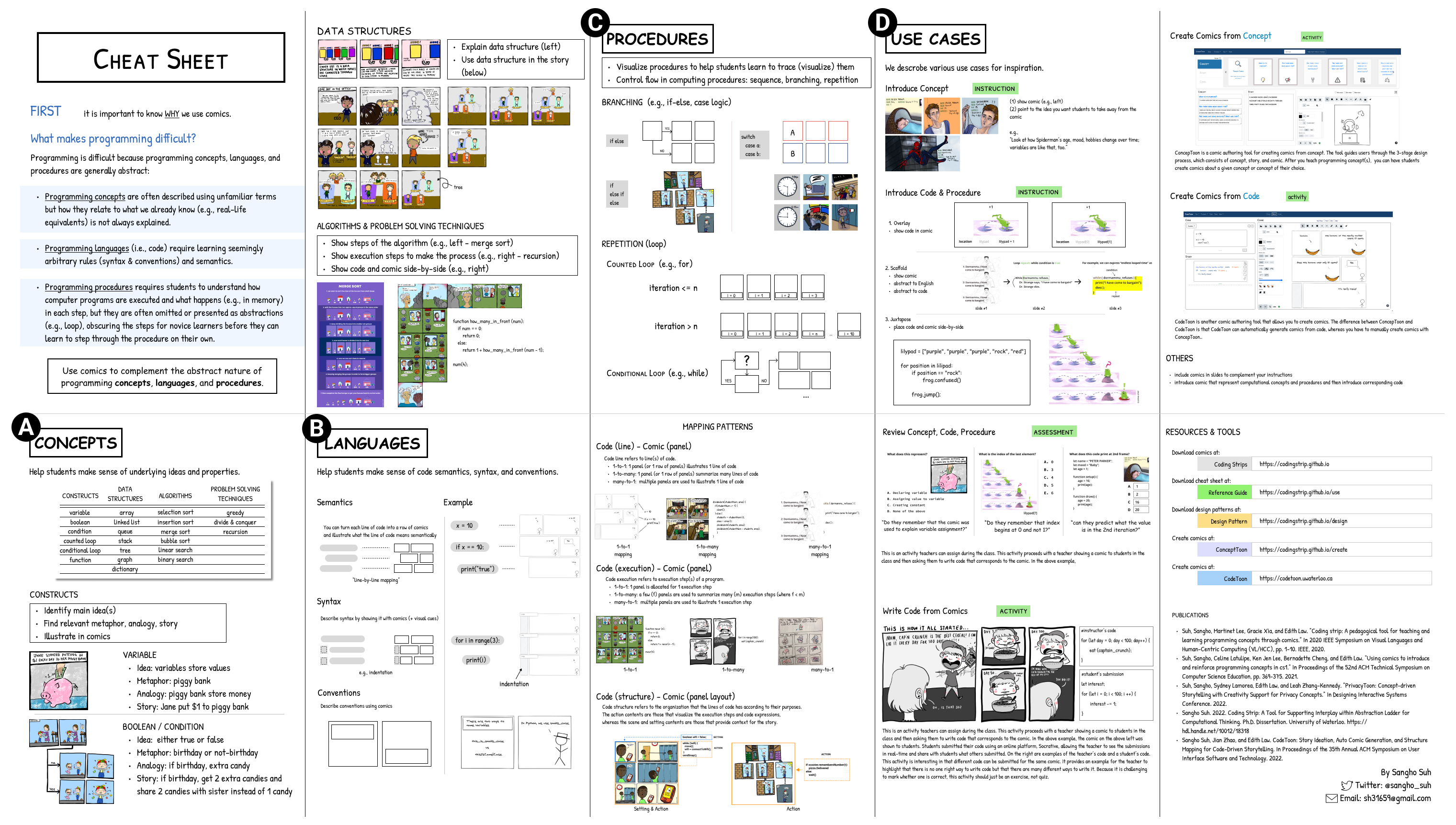}
  \caption{Cheat sheet for teaching programming with comics. It consists of four main sections: (A) concepts, (B) languages, (C) procedures, and (D) use cases. It can be downloaded at: \url{https://codingstrip.github.io/use}}
  \Description{Figure of cheat sheet.}
  \label{fig:teaser}
\end{teaserfigure}

\maketitle

\section{Introduction}

A cheat sheet is most commonly known as a set of well-structured, concise notes that summarize key concepts, techniques, or topics taught in a course that students may reference during their exams. While this is where the word originates from, the concise format of cheat sheet has evolved beyond the classroom setting and has become a popular format and term for documenting a variety of information---e.g., facts, tips, and techniques in a wide range of areas---in a concise manner~\cite{wang2020cheat}. For instance, cheat sheets are used to describe a set of commands for operating systems (e.g., Linux), programming languages (e.g., Python), and softwares, among many others, for users to reference quickly.

Inspired by the accessible format of cheat sheet, we developed a cheat sheet for teaching programming with comics in the hope that this makes comic-based pedagogy more accessible to teachers. Specifically, this work is motivated by the following observations: (1) research on teaching programming with comics has successfully explored various fundamental tasks and learning activities, showing that it is ready to introduce ways to improve and innovate current teaching practices~\cite{suh2021using, suh2022phd}; (2) many comic authoring tools (e.g.~\cite{suh2022privacytoon, suh2022codetoon})---both research and commercial tools (e.g., Pixton~\cite{pixton})---are now available, rendering the once feared task of creating comics no longer an obstacle for teachers; (3) the remaining friction to its adoption is the lack of clear, concise document that teachers can easily reference. Motivated by these observations, this work aimed to summarize comic-based pedagogy in an accessible cheat sheet format.


To develop this cheat sheet on teaching programming with comics, we referenced prior work on teaching programming with comics~\cite{suh2020coding, suh2021using, suh2022codetoon, suh2022phd}, analyzed 26 coding strips---a form of comic strips with its corresponding code---and their 30 design patterns~\cite{suh2020coding, codingstrip}, and formulated a concept-language-procedure framework to delineate and organize ways in which comics can support diverse needs in teaching programming.
For the preliminary evaluation, we presented it to 11 high school CS teachers at an annual conference for computer studies educators, administered a survey after this session, and analyzed their responses. Analysis of survey responses and feedback suggests that teachers find our cheat sheet easy to read/understand, useful, and well structured. In summary, our contributions include:

\begin{itemize}
    \item analysis of comics under the concept-language-procedure framework to delineate the values and benefits it offers;
    \item cheat sheet for teaching programming with comics that provides guidelines on why and how to use comics;
    \item evaluation of this cheat sheet by high school CS teachers.
\end{itemize}

\section{Related Work}

In this section, we first review prior work on teaching programming with comics. We then discuss what approaches have been explored to teach programming concepts, languages, and procedures and how comics can offer the same benefits and more.

\subsection{Teaching Programming with Comics}
Prior research on teaching programming with comics can be summarized as (1) using comic books (e.g.,~\cite{helloruby}) with characters presenting programming concepts via narration and (2) providing comics with their corresponding code (coding strip~\cite{suh2020coding}). Between the two, the coding strip research has been more active. Suh et al.~\cite{suh2020coding} developed and tested a design process and tools for creating coding strips via design workshops with students and teachers. Their team also tested four use cases for coding strips in an introductory CS course, where the use cases represented four basic teaching tasks for any programming classes~\cite{suh2021using}. Two tools have been developed to facilitate the authoring of coding strips: a comic authoring tool~\cite{suh2022privacytoon} based on the above-mentioned design process and tools~\cite{suh2020coding} and a visual programming environment that can automatically generate comics from code~\cite{suh2022leveraging}. Our cheat sheet summarizes these researches into a concise, accessible format of cheat sheet so that teachers can conveniently reference them without having to read lengthy research papers.


\subsection{Other Approaches: Storytelling, Animation, and Concrete Representations}

We assume that teaching programming boils down to teaching its concept, language, and procedure, and that they are all responsible for making programming abstract and less accessible for novice learners. We describe the approaches that have been used to support the teaching of programming concepts, languages, and procedures.

\begin{itemize}
    \item \textbf{Concept}: Storytelling~\cite{kelleher2007storytelling}, metaphor~\cite{colburn2008metaphor, esper2013codespells}, and analogy have been popular choices for making abstract programming concepts more intuitive and engaging. CS unplugged is a kinesthetic learning method in which students learn computing concepts through hands-on activities without the use of computers~\cite{nishida2009cs}. 
    \item \textbf{Language (Syntax \& Semantics)}: Turning abstract text-based programming into more concrete, graphic-based languages has been a popular approach. Graphic languages such as Scratch~\cite{resnick2009scratch} (a block-based programming language) provide programming constructs resembling LEGO blocks, making code expressions more accessible to younger audiences. Physical (tangible) programming makes programming more approachable, hands-on, and accessible to younger students and the visually impaired~\cite{mcnerney2004turtles, morrison2020torino}. A set of simplified text-based commands specific to a game environment has also been explored~\cite{lee2011personifying}.
    \item \textbf{Procedure}: Tracing execution steps in a program is a skill novice learners need to master, but the challenge is that in many cases all the steps are not necessarily shown and are instead presented as abstractions~\cite{suh2020promoting}. For instance, when describing a program, it is common to describe the number of times a program loops but omit what happens and how memory changes in each iteration. While this is to focus on the structural (high-level) aspects of the program, this makes procedures (sequence of steps) abstract for learners who have not yet mastered the ability to process the steps in their head. To address this, many have researched program visualization and developed new tools; they used animations of visual objects and models (e.g., characters, robot, and animals in Scratch, Gidget, Logo) as proxies to show how the code works; similarly, Storytelling Alice~\cite{cooper2000alice} illustrated procedures in terms of characters and stories; others (e.g., Python Tutor) have used interactive visualizations with diagrams, where learners can step through each line of code and observe changes in memory state, to understand the execution sequence and what each line of the code does to memory states~\cite{guo2013online}.
\end{itemize}

\subsection{Relevance of Comics}
When describing reasons to use comics, many frequently reference its ability to engage readers and improve memory~\cite{mccloud1993understanding, mayer2005cognitive, wang2019comparing}. Although they are true, we focus on other less-mentioned aspects of comics--(1) expressiveness and (2) affordance---that deserve to be mentioned as attributes as they are critical to making comics a powerful, appropriate medium for teaching programming.

\noindent\textbf{(1) Expressiveness.} Below, we highlight how the visual language of comics can illustrate abstract concepts, languages, and procedures in programming in effective and creative ways.

\begin{figure}[h]
	\centering
	\includegraphics[width=0.45\textwidth]{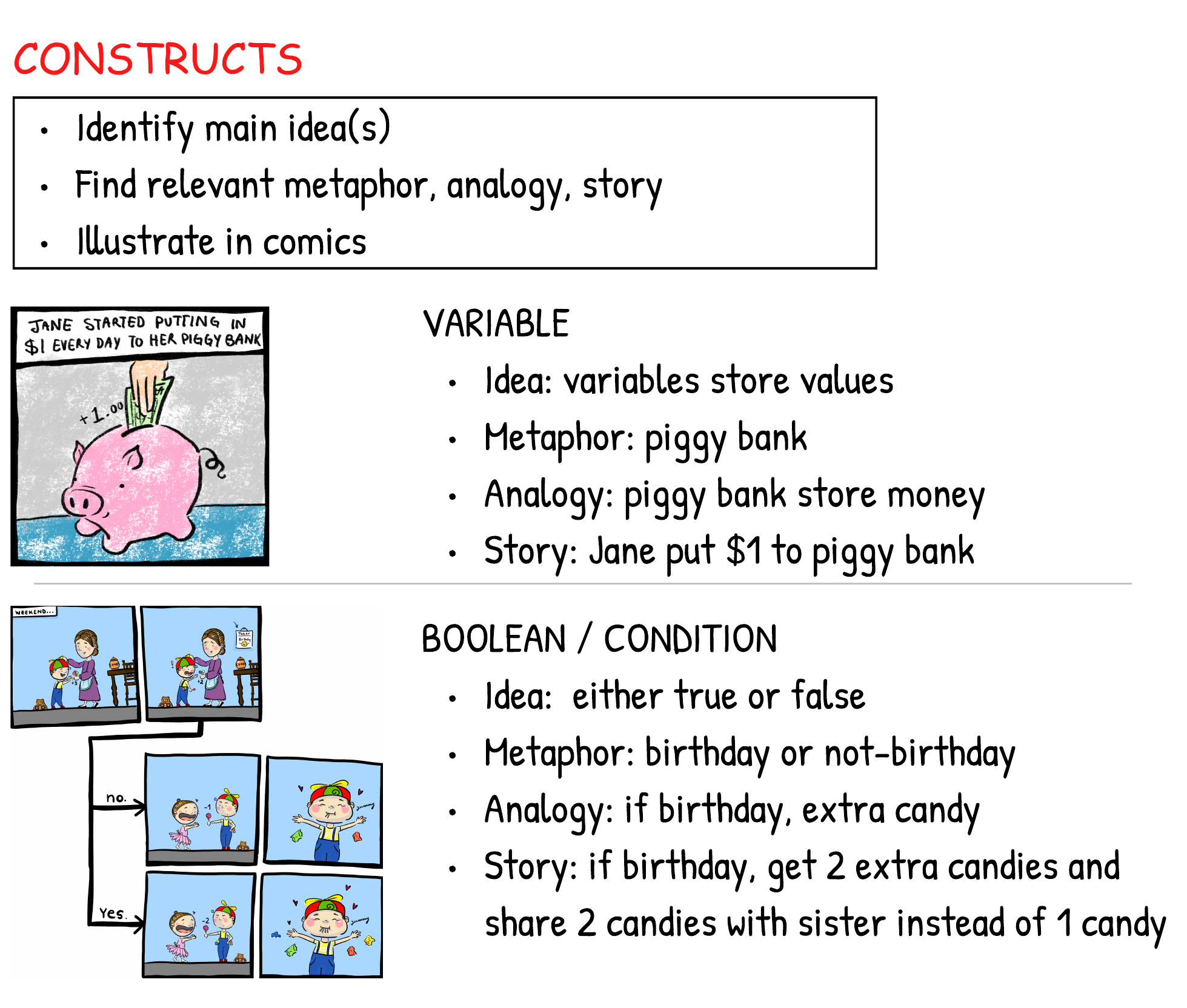}
	\caption{Part of the concept section (Fig.~\ref{fig:teaser}A)}
	\label{fig:concept-section}
\end{figure}

\begin{figure}[h]
	\centering
	\includegraphics[width=0.45\textwidth]{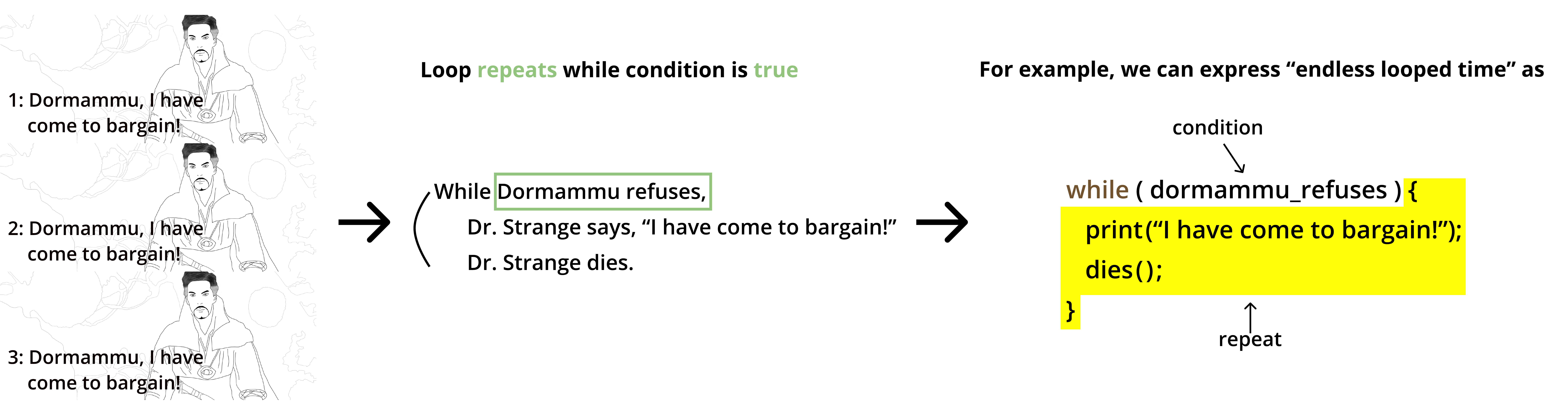}
	\caption{An example of how comics can be used to teach code semantics \& syntax~\cite{suh2021using}}
	\label{fig:comic-to-code_semantics}
\end{figure}

\begin{figure}[h]
	\centering
	\includegraphics[width=0.45\textwidth]{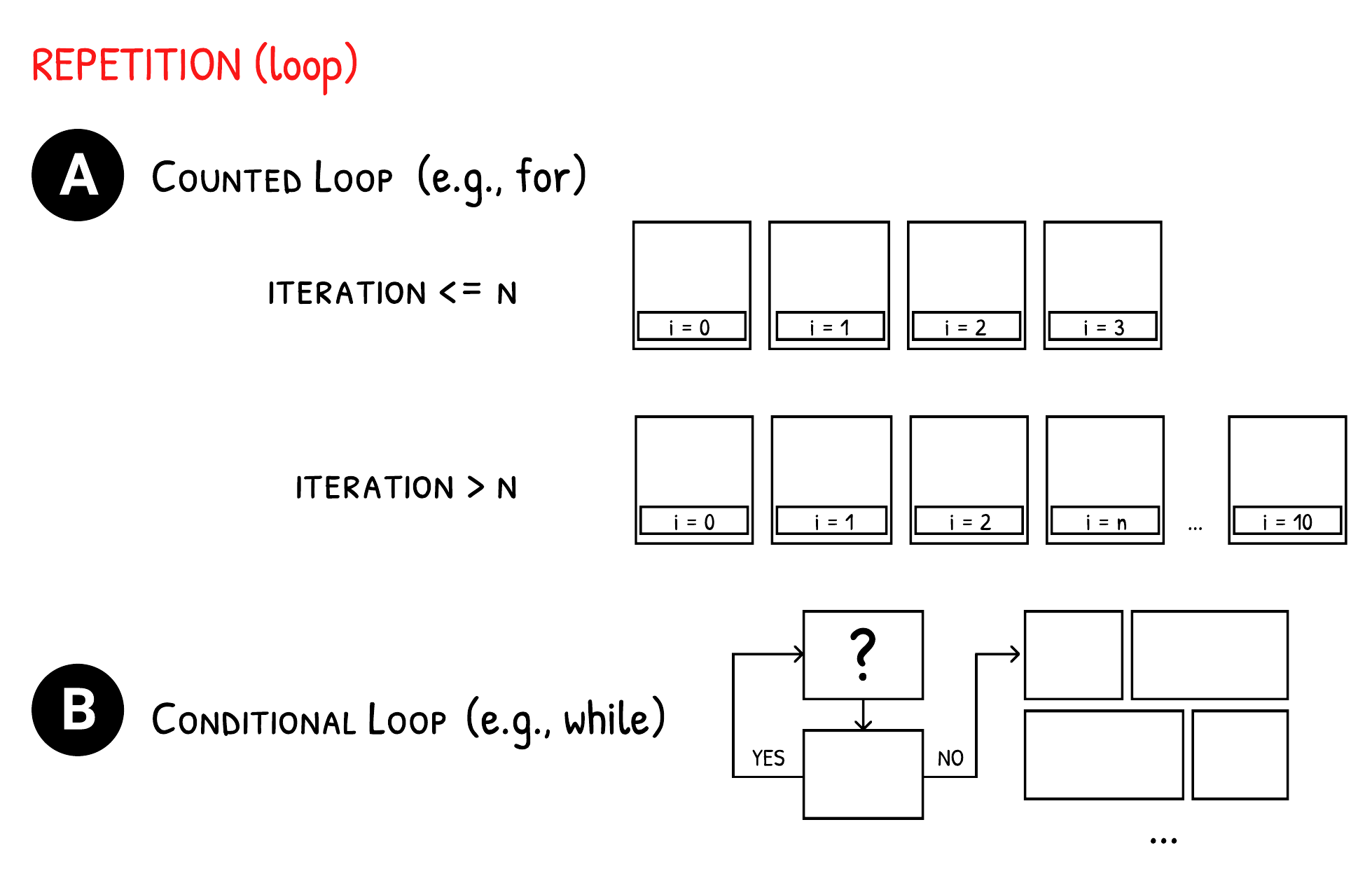}
	\caption{Part of the procedure section (Fig.~\ref{fig:teaser}C)}
	\label{fig:procedure-section}
\end{figure}

\begin{itemize}
    \item\textbf{Concept}: Comics is a unique medium that can leverage the power of storytelling, metaphors, and analogies. Fig.~\ref{fig:concept-section} shows Jane putting \$1 to her piggy bank. Here, piggy bank is a metaphor for variable that stores something (\$1 in this case). Jane putting money into it is analogous to a value being assigned to a variable in semantics. On top of storytelling, metaphor, and analogy, comics adds visual representations to the mix. Thus, it can arguably provide (or leverage) the best of both worlds, offering cognitive benefits visual representations possess. Furthermore, it can also address the limitations of metaphors and analogies, as comics can leverage visuals to constrain interpretations and minimize the risk of confusion often cited as limitations of metaphors and analogies.
    
    \item\textbf{Language (Syntax \& Semantics)}: Code syntax and semantics can be illustrated and explained by using comics in a creative way. For instance, Fig.~\ref{fig:comic-to-code_semantics} shows how comics can be used to scaffold a piece of code. Students see a comic (visual language) about the movie \textit{Dr. Strange} where a character uses the Time Stone to start a time loop to repeat the same moment; then students see this expressed in English (natural language), and then in code (programming language). The example shows how comics can be used to teach the language syntax and semantics: the indented structure carries over from comic to English and then to code to teach this syntax; the repetition of the same panels provides visual clarity for users to grasp what the programming construct \code{while} does; comics can help users better understand and retain the idea longer by leveraging a familiar story and visuals; the code semantics for \code{print()} is also explained with the action of a character speaking, providing intuition by associating it with a familiar abstraction.
    \item\textbf{Procedure}: Also referred to as ``sequential art,'' one of the defining characteristics of comics is its sequential nature. It can leverage sequence of panels to illustrate computational procedures in creative ways. Fig.~\ref{fig:procedure-section} shows two comic examples on loop. The two examples represent two slightly different ways of viewing (and visualizing) loops. Fig.~\ref{fig:procedure-section}A illustrates counted loop (e.g., \code{for}), visualizing a loop with fixed number of iterations. Fig.~\ref{fig:procedure-section}B illustrates conditional loop (e.g., \code{while}), where the number of iterations is not pre-determined. 
\end{itemize}

\begin{figure}[h]
    \centering
    \begin{subfigure}[t]{0.22\textwidth}
        \includegraphics[trim=0cm 0cm 0cm 0cm, clip=true, width=\textwidth]{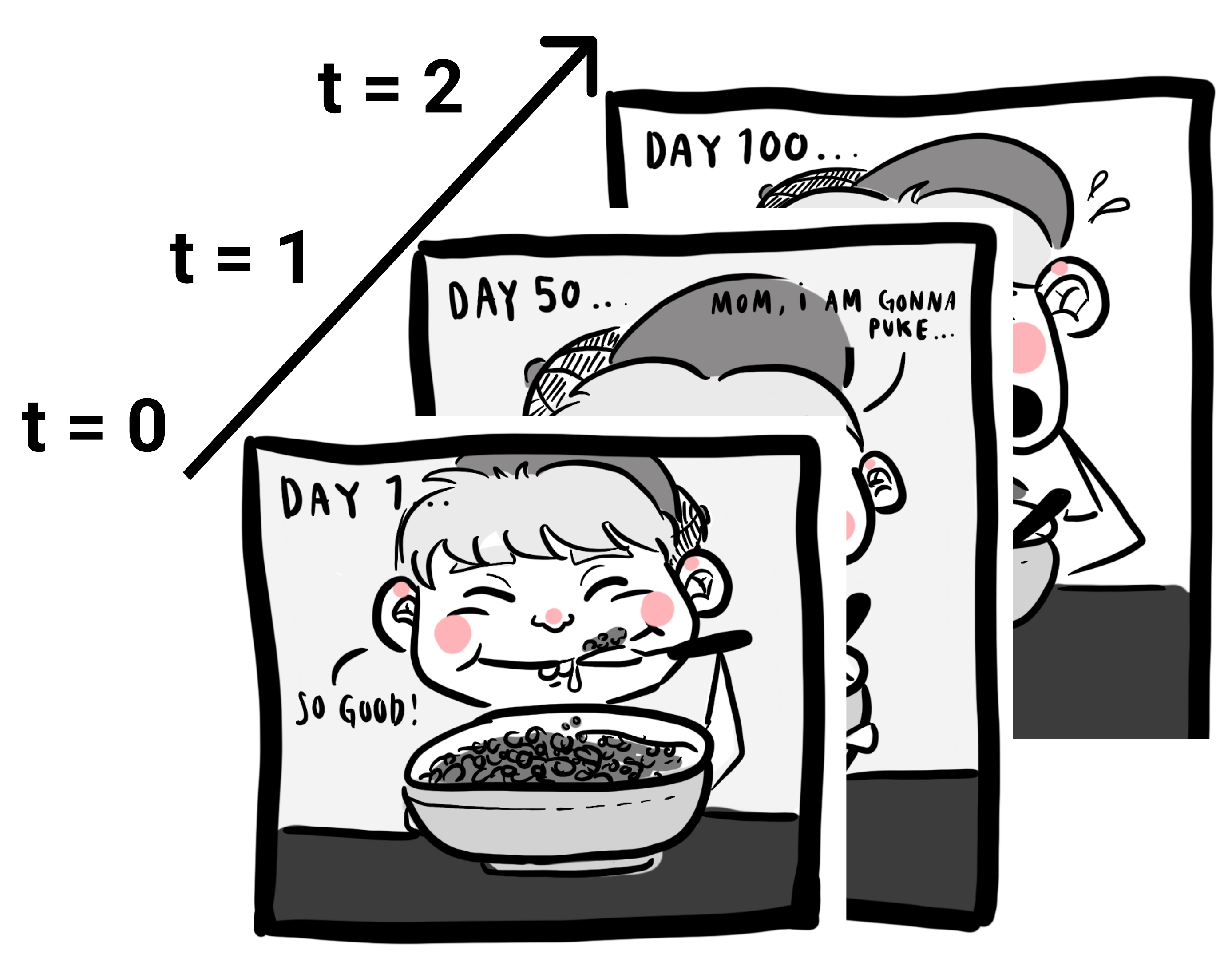}
        \caption{Animation}
        \label{fig:animation-vs-comic}
    \end{subfigure}
    \begin{subfigure}[t]{0.22\textwidth}
        \includegraphics[trim=0cm 0cm 0cm 0cm, clip=true, width=\textwidth]{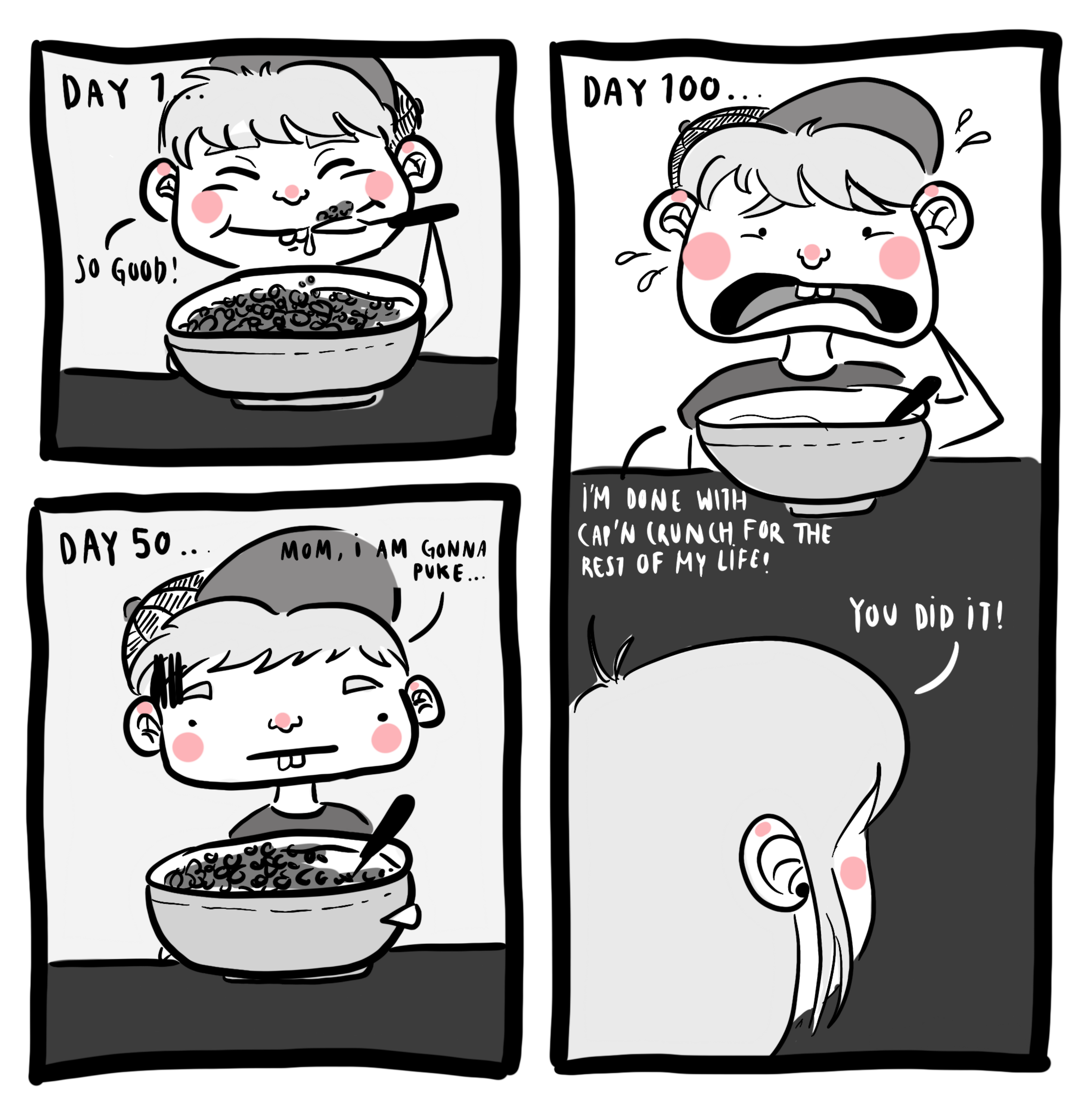}
        \caption{Comic}
        \label{fig:comic-vs-animation}
    \end{subfigure} 
    \caption{While animation (a) shows one scene at a time, comic (b) shows multiple scenes at once and provides readers greater sense of control over the pace at which they process the information~\cite{yang2016comicsbelong}. Furthermore, being able to see all the scenes altogether allows one to discover any relationship (e.g., connection between the scenes/states) more easily.}
    \label{fig:comic-vs-animation-comparison}
\end{figure}

\noindent\textbf{(2) Affordance.} Comics provides unique cognitive benefits that cannot be found in similar media such as animation. As shown in Fig.~\ref{fig:comic-vs-animation-comparison}, comics differs from animation in that sequence of events can be juxtaposed and shown all at once using panels---not to mention various ways in which panels can be arranged in creative ways to enhance the delivery and reading experience~\cite{cohn2015navigating, bach2018design}. While the dynamic nature of animation may be more effective in terms of engaging viewers, comics offers unique benefits in some ways, e.g., placing multiple panels side-by-side to help viewers easily compare and contrast events occurring at different times and perceive connections between separate scenes. Also, Yang~\cite{yang2016comicsbelong} found that his students who received his math instructions in the form of both comics and video preferred comics: with video, students had to constantly rewind it to re-watch the parts they did not understand; with comics, they could read at ``their pace,'' which empowered them as they do not have to re-wind and be reminded again about the parts they did not understand.
Jonathan Hennessey explains the unique affordance of comics this way~\cite{web:teaching-content-with-comics}:

\begin{quote}
    ``In one image or composition, the reader can linger over the potential significance of small details without having the sense that the narrative flow has been disrupted. No matter how much time you spend inside one panel, you never feel like the story has stopped or altered tempo. Not the way you would if you pressed `pause' while watching a video. So it's ideal for students and teachers. It has the vividness of the moving image and the complexity of text.''
\end{quote}

The unique affordance of comics can be useful in the context of teaching programming. In programming classes, novice learners often vary in their ability to keep up with the teaching. Being able to control the \textit{tempo} can be useful for students who have difficulties keeping up with the instructions in class. It is useful in terms of being able to accommodate students at different levels. For example, struggling students can``linger over'' the initial panels longer and other students who do not need to do so can pay attention to the later part of the sequence. Students can read different parts of the comics and not have to worry about interfering with the pace at which other students are progressing, because the sequence of panels can all be displayed, as shown in Fig.~\ref{fig:comic-vs-animation-comparison}. Further, when an instructor takes a piece of code and steps through the execution sequence, some students can get confused about a certain explanation or step but have difficulty specifying which part or step they were confused about; with comic, students can easily pinpoint which part they were confused about.



\section{Cheat Sheet: Design Goals}

In this section, we describe considerations and design goals (\textbf{DGs}) that guided the design of our cheat sheet:

\textbf{DG1. Structure cheat sheet to support fundamental tasks in lesson planning}. To situate our cheat sheet in a setting where teachers can use them during their lesson planning, we reviewed lesson plans and resources teachers use to construct their lesson plans.
As we reviewed a number of lesson plans found in Creative Computing Curriculum~\cite{creativecomputing}, K–12 Computer Science Framework~\cite{k2016k}, and others, we found that they can have different components depending on the needs of the classroom~\cite{monett2015evolving}. Thus, to make our cheat sheet generalizable, we chose to focus on three fundamental tasks in lesson planning: (1) defining learning objectives, (2) crafting learning activities/instructions, and (3) identifying assessment methods. This motivated us to include instructions such as ``identify main idea(s)'' (Fig.~\ref{fig:concept-section}) and use cases related to reviewing the learned concepts, languages, and procedures.

\textbf{DG2. Make cheat sheet flexible to accommodate diverse age groups and programming paradigms.} Prior research suggests that the target student groups for this comic-based pedagogy do not need to be limited to a younger audience (e.g., K-5)~\cite{suh2021using, suh2022phd}. In other words, comics can be used in programming classes for K-5 with block-based programming~\cite{suh2022codetoon} as well as for university-level introductory programming classes taught using text-based programming languages (e.g., Python). Thus our cheat sheet should not center around specific programming languages or paradigms but should be organized using a higher level of abstraction. This led us to search and specify concept, language, and procedure as basic programming components that students are taught regardless of their age and programming language. This is why we structured our cheat sheet around concept, language, and procedure.

\textbf{DG3. Explain why they might want to use comics.} Unlike some cheat sheets that list a set of commands or shortcuts to help people with `how,' our cheat sheet needs to go beyond simply helping with `how.' This is because this cheat sheet aims to help teachers make informed decisions on whether to use comics and, if yes, which one they should choose. Thus, our cheat sheet explains `why' to help its users make informed decisions. This made us include at the beginning of our cheat sheet `what makes programming difficult' in order for us to explain why they might want to use comics. Also, this is why---in the instruction for the procedure section in the cheat sheet---we write: `Visualize procedures to help students learn to trace (visualize) them.'

\textbf{DG4. Provide examples and use cases to make the content and ideas clear.} While surveying existing cheat sheets, we found that even for cheat sheets on the same topic, they could differ in terms of their focus. For instance, one version of the cheat sheet for Python lists only a set of built-in functions, while another version lists explanations on basic syntax and conventions. The difference stems from which group of users the cheat sheet is for. Since our target group is teachers with no prior knowledge of teaching programming with comics, we need to include examples and use cases to make the content and ideas clear for them.

\section{Preliminary Evaluation}

To evaluate our cheat sheet, we presented it at a conference for computer science educators, an annual conference organized by the University of Waterloo for high school CS teachers. For this year, the conference was held virtually.

Participants were informed through the abstract in the conference schedule that a survey will be administered during our session but that they are free to attend the session and not partake in it. The 50-minute session proceeded with the presenter (the author of this paper) first presenting the cheat sheet (40 min) and then waiting for participants to answer the survey (10 min).

The survey consisted of 5-point Likert scale questions that probed their perceptions of the cheat sheet's readability, usefulness, and organization. Questions with options to express the level of agreement (e.g., Somewhat Agree) were bipolar (extremely negative to extremely positive),\footnote{Strongly Disagree, Somewhat Disagree, Neither Agree nor Disagree, Somewhat Agree, Strongly Agree} whereas those for assessing the usefulness were unipolar (0 to extreme).\footnote{Not At All Useful, Slightly Useful, Moderately Useful, Very Useful, Extremely Useful} Each Likert scale question was followed by a question asking participants to elaborate on their responses (i.e., qualitative feedback). The survey ended with demographic questions about their age, gender, and teaching experience.

Seven out of 11 attendees (age: M=51.4, SD=14.4; gender: 1F, 6M) answered our anonymous survey. Except for one participant who had no teaching experience, all of them were highly experienced in teaching (M=20, SD=7.5); they had equally many years of experience teaching computer science (M=17.3, SD=10.5; Range: [6, 34]) for students in grades 10 - 12 (4 Grade 10; 5 Grade 11; 4 Grade 12). The programming languages they taught included: Turbo Pascal, Visual Basic, BASIC C, C++, C\#, Java, Scratch, Python, Processing, Java/Swift. We refer to these participants as {P1}\dots{P7}.

\textbf{Readability.} 
Participants generally found the cheat sheet (1) easy to read (1 Strongly Agree, 6 Somewhat Agree), (2) easy to understand (7 Somewhat Agree), and (3) well-structured (2 Strongly Agree, 5 Somewhat Agree). Several comments were related to problems caused by the set up of the session. P3 said, ``\textit{the cheat sheet was small on the screen}.'' P5 mentioned that it was a bit of challenge to ``\textit{grasp concepts}'' since he did not have ``\textit{hands on access' to the cheat sheet and it was a ``short session}.'' P4, on the other hand, commented on the content, saying that ``\textit{almost too much information is shown}'' and that he is ``\textit{most interested in the basics}.'' 

\textbf{Usefulness of Comics.}
While they generally had positive attitudes towards comics (`I like comics': 3 Strongly Agree; 2 Somewhat Agree; 2 Neither Agree nor Disagree), surprisingly many of them---prior to the session---did not think comics could be useful for teaching/learning programming (3 Not At All Useful, 2 Slightly Useful, 2 Moderately Useful). All of them, however, did indicate that---after the session (during which they learned various ways comics can be used to teach programming through the cheat sheet)---their perception towards its usefulness became more positive (7 answered More Useful after This Session to `Compared to before this session, how useful do you think using comics to teach/learn programming is?'). The participants mentioned several reasons for their responses, explaining that (1) comics seem to make programming ``\textit{more accessible}'' (P1) and can be useful, especially for students in ``\textit{ESL/ELL and IEPs}'' (P3). P5 said:

\begin{quote}
    ``\textit{This could be supplemental to what we're already doing and a great addition to the course.  It will really help some students who like learning with visuals}.''
\end{quote}

\textbf{Usefulness of Cheat Sheet.} 
Participants mostly found the cheat sheet very useful (3 Extremely Useful, 2 Very Useful, 2 Moderately Useful). They seemed to agree that it is very ``\textit{detailed}'' and contains a ``\textit{clear source of information}'' for understanding how to teach with comics. P3 saw it as a ``\textit{great place to start, [as it offers] concrete examples}.'' P4 said: ``\textit{I can see students being self-motivated to try this out and easily pick up these concepts}.'' On the other hand, P7 who found it Moderately Useful explained that it seems to ``\textit{require some explanation}'' before one can use it.

\textbf{Interest in Using Cheat Sheet.} 
About half of the participants (3 Probably Yes, 4 Might or Might Not) expressed interest in using it to teach programming. P5 had an interesting idea of sharing this cheat sheet with his students for them to read, saying ``\textit{I would make this available to students after explaining it so they can try it out as an assignment or an experiment}.'' Two participants who were on the fence explained that they just ``\textit{need time to process how to incorporate this into [their] teaching practice}'' (P4) and ``\textit{would like to take a closer look at the cheat sheet to better evaluate the benefit}'' (P7).

\textbf{Concept-Language-Procedure Framework.} 
Participants generally agreed that programming can be thought of as consisting of concept, language, and procedure (3 Strongly Agree, 4 Somewhat Agree). P4 said: ``\textit{this helped me organize my thoughts}.'' P7 also approved, saying: ``\textit{I hadn't considered structuring coding in this way, but it makes sense. It is one logical way you could do it}.'' They were also generally in agreement with how the cheat sheet was structured around concept, language, and procedure (3 Strongly Agree, 1 Somewhat Agree, 3 Neither Agree nor Disagree).

\section{Discussion}

\textbf{Opportunities.} With the growing popularity of CS education, the demand for a larger teacher workforce has increased, with an increasing number of research, resources, and workshops to serve this need. Our cheat sheet can be used as a teaching resource for teacher education. As comics is a popular medium for students, incorporating comics into programming classes can be a welcome innovation in classrooms, as prior work has shown~\cite{suh2021using}. Besides programming, other more advanced CS classes---such as operating system and network---can also benefit from leveraging comics. It would be interesting to explore, e.g., what kinds of design patterns can be used to teach concepts in other areas in CS.

\textbf{Limitations.} As this cheat sheet summarizes relatively recent research on teaching programming with comics, we do not see---nor present---this cheat sheet as a final and comprehensive summary of teaching programming with comics. As new techniques and tools are developed in the future, this cheat sheet should be updated. The contributions of this work are in initiating the first step towards summarizing the comic-based pedagogy and formulating the concept-language-procedure framework to place it inside.

Even though our cheat sheet seems to have done a good job at compressing years of research on teaching programming with comics, one participant's comment that explanation is still required to understand how to use the cheat sheet shows that a walk-through video or article explaining the cheat sheet might still be necessary. But as noted in our design goals (\textbf{DG3} \& \textbf{DG4}), this may be inevitable as long as the cheat sheet is used by teachers without prior experience or knowledge of teaching programming with comics.

Moreover, our evaluation did not extend to verify whether teachers can use this cheat sheet to successfully incorporate the comic-based pedagogy into their teaching. As a result, this work does not provide a complete picture of what, if any, worked and did not.


\textbf{Future Work.} As highlighted above, we need to test whether teachers can successfully plan their lessons using our cheat sheet, to understand, for example, how much time and effort are required to adopt this approach, as well as what improvements, if any, are needed to make the adoption easier. Moreover, while the presented cheat sheet focuses on programming, comics can also be used to teach other CS courses, such as computer networks, privacy (cf.~\cite{suh2022privacytoon}), ethics, and AI. We plan to explore how comics can be used to teach other topics and areas in computer science and share a collection of their cheat sheets in \url{www.codingstrip.github.io}. The teachers in our study also asked for a cheat sheet with examples of comics for all basic programming concepts. We believe access to such a resource will make using comics easier and attract interest from teachers. Another interesting idea would be to explore the possibility of using cheat sheet as a learning resource for students. As P5 mentioned above, cheat sheet can be very helpful to students. For example, prior research has shown that the process of creating a cheat sheet helps students by forcing them to organize the learned concepts~\cite{de2012student}. It may be interesting to see whether having students create a cheat sheet for programming with comics helps them learn better.
Finally, a recent work showed that we can use comics to elicit discussions around how AI models are trained among those without training in or knowledge of AI~\cite{kuo2023understanding}. We could potentially use this as an opportunity for students to learn AI ethics.

\section{Conclusion}
In this work, we developed a cheat sheet for teaching programming with comics. To develop this cheat sheet, we reviewed prior work on this topic, analyzed 26 coding strips, and formulated the concept-language-procedure framework to delineate various ways comics can be used to facilitate the teaching of programming concepts, languages, and procedures. We evaluated this cheat sheet with seven high school CS teachers and found that they find it easy to read/understand, useful, well-structured, and are interested in learning more about teaching programming with comics. 

\begin{acks}
This research was funded by the Learning Innovation and Technology Enhancement (LITE) Seed Grant at the University of Waterloo. I would like to thank Sandy Graham for the invitation to present at the CEMC Bringing Teachers Together Virtually Conference and teachers for their participation and feedback.
\end{acks}

\bibliographystyle{ACM-Reference-Format}
\bibliography{main}

\appendix

\end{document}